\documentclass[10pt]{article}

\usepackage[authoryear]{natbib}

\usepackage[T1]{fontenc}
\usepackage[utf8x]{inputenc}
\usepackage[english]{babel}

\usepackage{graphicx}
\usepackage[allcolors=black,colorlinks=true]{hyperref}

\usepackage{times}

\title{Automatic Identification of Research Fields in Scientific Papers}

\usepackage{authblk}
\author[1]{Eric Kergosien}
\author[5]{Amin Farvardin}
\author[3]{Maguelonne Teisseire}
\author[2]{Marie-No\"elle Bessagnet}
\author[1,6]{Joachim Sch\"opfel}
\author[1]{St\'ephane Chaudiron}
\author[1]{Bernard Jacquemin}
\author[2]{Annig Lacayrelle}
\author[3,4]{Mathieu Roche}
\author[2]{Christian Sallaberry}
\author[3,4]{Jean Philippe Tonneau}

\affil[1]{Univ. Lille, EA 4073 \textendash{} GERiiCO, F-59000 Lille, France\\prenom.nom@univ-lille.fr}
\affil[2]{LIUPPA, Universit\'e de Pau et des Pays de l'Adour, Pau, France\\ prenom.nom@univ-pau.fr}
\affil[3]{TETIS, Univ. Montpellier, APT, Cirad, CNRS, Irstea,  Montpellier, France\\prenom.nom@teledetection.fr}
\affil[4]{Cirad, Montpellier, France\\prenom.nom@cirad.fr}
\affil[5]{LAMSADE, Universit\'e Paris-Dauphine, Paris, France\\MohammadAmin.Farvardin@dauphine.eu}
\affil[6]{ANRT, Lille, France\\Joachim.Schopfel@univ-lille3.fr}

\date{\vspace{-3ex}}

\begin{document}

\maketitle

\noindent\textbf{Abstract:} The TERRE-ISTEX project aims to identify scientific research dealing with specific geographical territories areas based on heterogeneous digital content available in scientific papers. The project is divided into three main work packages: (1) identification of the periods and places of empirical studies, and which reflect the publications resulting from the analyzed text samples, (2) identification of the themes which appear in these documents, and (3) development of a web-based geographical information retrieval tool (GIR). The first two actions combine Natural Language Processing patterns with text mining methods. The integration of the spatial, thematic and temporal dimensions in a GIR contributes to a better understanding of what kind of research has been carried out, of its topics and its geographical and historical coverage. Another originality of the TERRE-ISTEX project is the heterogeneous character of the corpus, including PhD theses and scientific articles from the ISTEX digital libraries and the CIRAD research center. \\[1ex] 

\section{Introduction}

Widespread access to digital resources, via academic platforms \textendash{} for example, the Gallica project (BnF)\footnote{\url{http://gallica.bnf.fr/}}, the ISTEX\footnote{\url{http://www.istex.fr/}} platform, electronic theses and dissertation repositories (TEL), content federation services (Isidore), or electronic publishing platforms (OpenEdition) \textendash{} offers numerous possibilities for users. The ISTEX initiative was launched to create innovative information retrieval services and provide access to digital resources through different search processes. The increasing adoption of information and communication technologies by researchers in different academic disciplines, especially in the social sciences and humanities, is changing the conditions of knowledge appropriation. Digital humanities have made it possible to develop platforms, providing researchers with large volumes of academic papers and with support services to add value and make use of them (e.g., the TELMA\footnote{\url{http://www.cn-telma.fr/}} application).

The TERRE-ISTEX project was developed in this research context and proposes (1) to identify the covered territories and areas from scientific papers available in digital versions within and outside the ISTEX library, and (2) to evaluate the academic disciplines involved (e.g. history, geography, information sciences, etc.) as well as the evolution of disciplinary and multi-disciplinary research paradigms in selected topics. The results of this project will help scientists working on a given territory (areas at different scales, such as township, region, country, or continent) to retrieve papers on the same territory.

\section{The TERRE-ISTEX Project}

\begin{figure}[hbt]
 \centering
 \includegraphics[width=.475\textwidth]{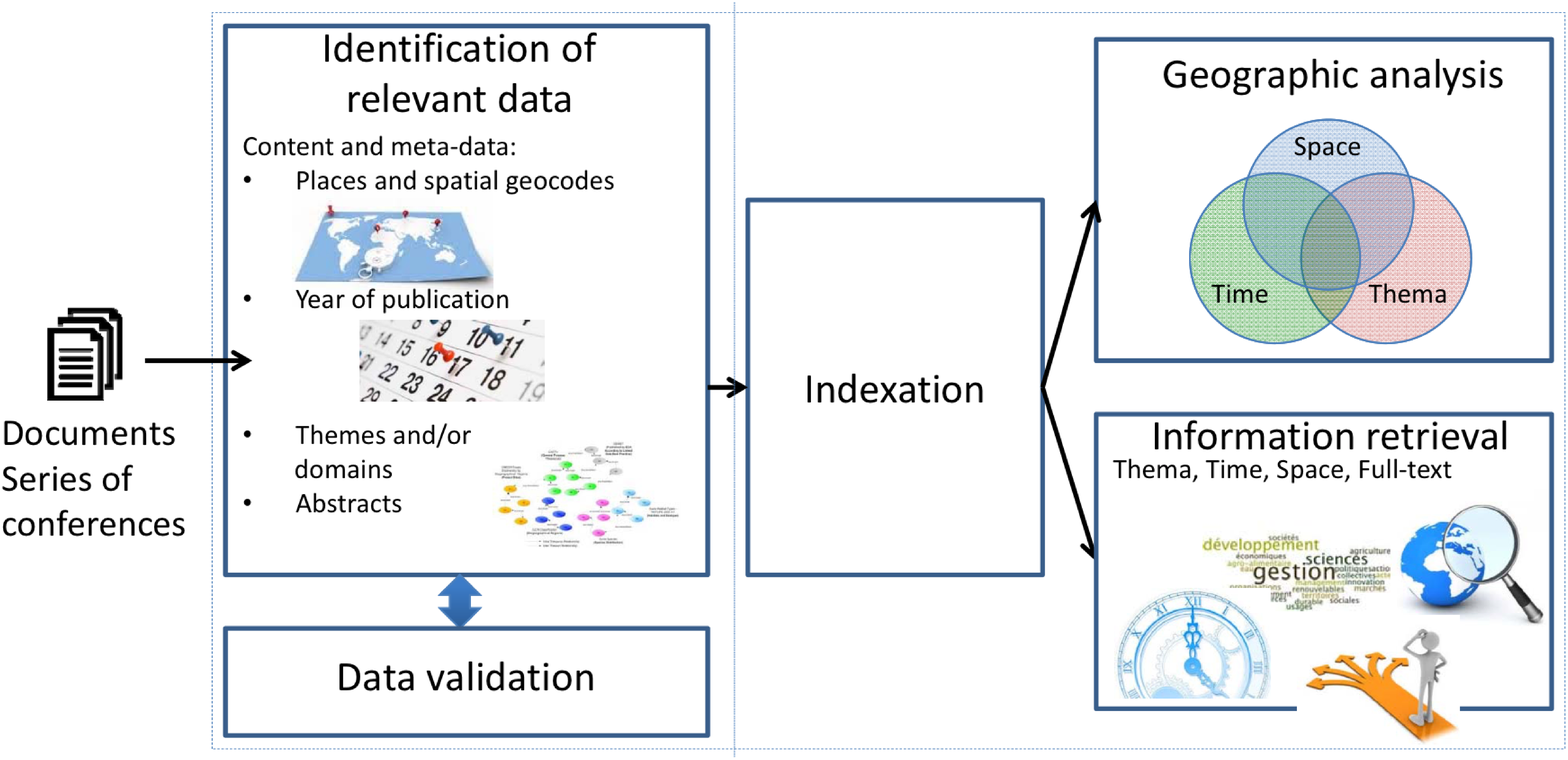}
 \caption{Generic Approach to Multidimensional Analysis of Scientific Corpus}
 \label{fig:genericApproach}
\end{figure}

The generic approach used in the TERRE-ISTEX project is described in Figure~\ref{fig:genericApproach}. Regardless of any scientific publications corpus, a first step is to standardize textual documents. A second step is to identify, in metadata and contents of documents, the research fields as well as the scientific disciplines involved. The research field is defined as the locations constituting the territory in which the research is conducted on a given date or period of time.

In the experimental section, we present a Web application called SISO to help domain experts to analyze and to validate annotated text samples. The indexed and validated data are then integrated into a documentary database in order to allow, on the one hand, the analysis of data and, on the other hand, the retrieval of papers on the same field and/or the same period and/or the same discipline or sub-domain using a Web-demonstrator of geographical information search.

The next section describes the corpus used within the framework of our project.

\subsection{The corpus}

Elaborating a corpus is a major prerequisite in the process of analysis and information retrieval. We targeted three sources of scientific papers, namely the ISTEX\footnote{\url{http://www.istex.fr/category/plateforme/}} platform, the Agritrop\footnote{\url{https://agritrop.cirad.fr/}} open archive (CIRAD\footnote{\url{http://www.cirad.fr}}), and a sample of PhD theses from the ANRT\footnote{\url{https://anrt.univ-lille3.fr/}} with associated metadata available on the portal \url{theses.fr}.

We conducted a case study on the topic of climate change in Senegal and Madagascar. We collected an initial corpus of documents from the ISTEX platform (about 170,000 documents) using queries with the following keywords: "climate change", "changement climatique", "Senegal", "S\'en\'egal", "Madagascar". From the same keywords, we collected 400 theses from the ANRT database. Finally, the documents from Agritrop are related to studies dealing with Madagascar and the Senegal River. The 92,000 references and 25,000 full-text documents include different types of academic papers, i.e., scientific publications (i.e., articles, etc.), grey literature (e.g. reports, etc.), and technical documentation. Each document is associated with metadata, including an abstract.

The metadata formats of the different items depend on the document origin: MODS\footnote{\url{http://www.bnf.fr/fr/professionnels/f\_mods/s.mods\_presentation.html}} (ISTEX), XML based on the Dublin Core (CIRAD), and RDF (ANRT). The corpus is multilingual: most of the documents are either in French or in English, but there are also documents using both languages (for example, with a summary in French and a summary in English). The corpus is thus composed of multilingual and heterogeneous documents, both in terms of content and format.

\subsection{An important step to standardize data}

\subsubsection{The process TERRE-ISTEX}

Initially, the process developed by the TERRE-ISTEX project is applied to metadata and abstracts. Because of the data heterogeneity, we chose to standardize metadata using the pivotal MODS (Metadata Object Description Schema) format, recommended on the ISTEX platform. The MODS format has several advantages: (a) it is suitable for describing any type of document and any medium (digital or print); (b) it is richer than the Dublin Core; and (c) it is similar to the models for structuring bibliographic information used in libraries (e.g. MARC format). For these reasons, we apply, in a first step, an algorithm of model transformation to those 92,400 documents of the corpus which do not comply with this format. The second step concerns the annotation in the abstracts of spatial, temporal, and thematic entities. This step is detailed below. As a result, the MODS-TI data model expands the MODS format to describe spatial, temporal, and thematic entities extracted from documents. The MODS-TI model is detailed in the following section. Step 3 implements a new algorithm for transforming the MODS-TI format in order to create indexes so that all data can then be processed in the final stages of analysis and information retrieval.

\subsubsection{The TERRE-ISTEX data model}

The TERRE-ISTEX data model expands the MODS format to describe spatial, temporal, and thematic information extracted from documents and from corresponding metadata. The choice of MODS was determined by the fact that MODS is the main format on the ISTEX platform and by the advantages described above.

\begin{figure*}[t]
 \centering
 \includegraphics[width=.95\textwidth]{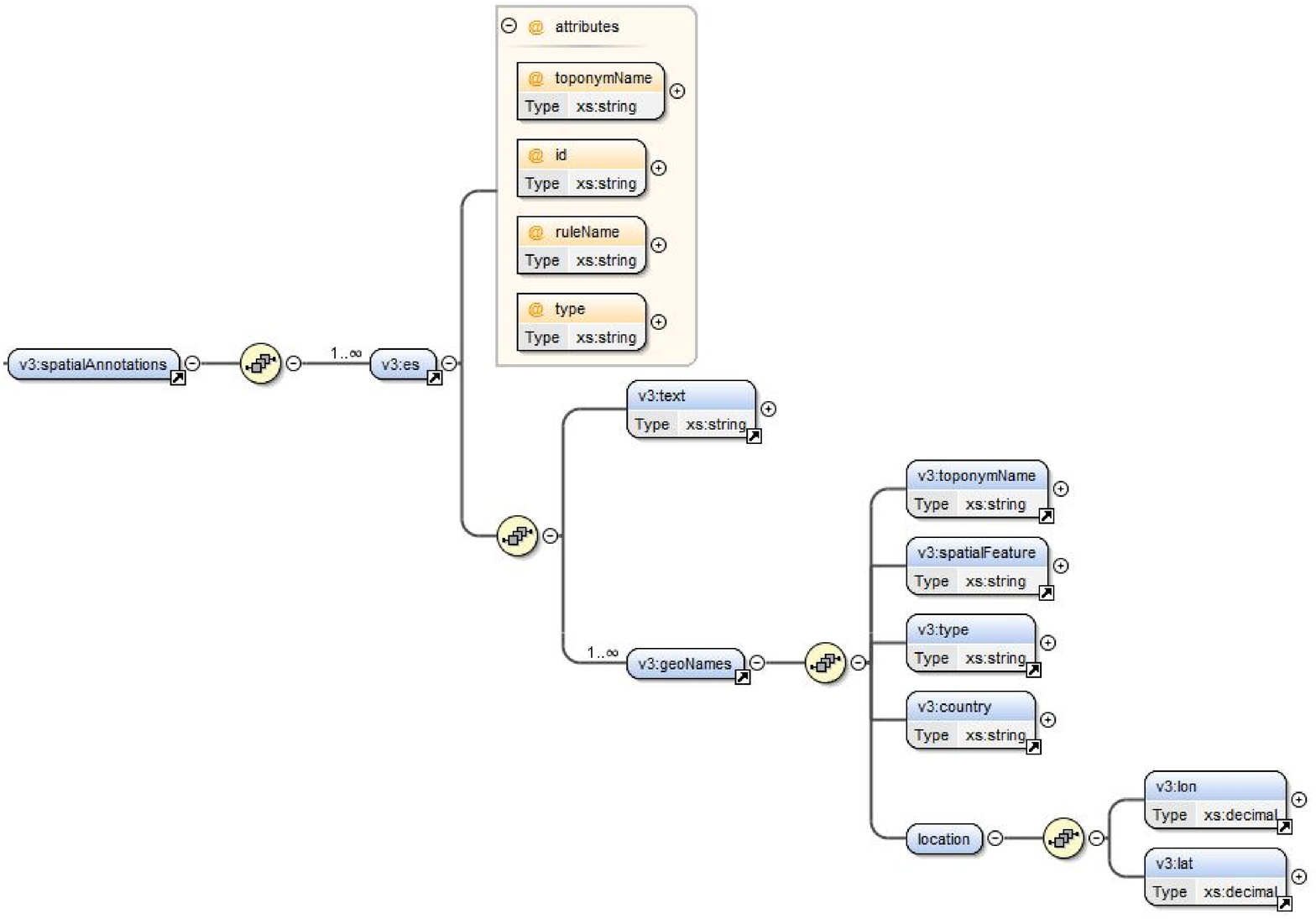}
 \caption{DTD describing the tag \texttt{<spatialAnnotations>}}
 \label{fig:spatialAnnotations}
\end{figure*}

We added three tag sub-trees to a MODS document:
\begin{itemize}
 \item \verb|<spatialAnnotations>|,
 \item \verb|<temporalAnnotations>|,
 \item \verb|<thematicAnnotations>|.
\end{itemize}
In the following, we give an example of the sub-tree for spatial entities (ES). The tag \verb|<spatialAnnotations>| contains a set of spatial entities (tag \verb|<es>|), with the annotated text for each of them. (tag \verb|<text>|) as well as its spatial footprint obtained by querying the GeoNames resource. The corresponding DTD is shown in Figure~\ref{fig:spatialAnnotations}.

\subsection{Named Entity annotation }

\subsubsection{Spatial entities}

In the TERRE-ISTEX project, the methodology is based on linguistic patterns for the automatic identification of spatial entities (ES) \cite{tahrat_text2geo:_2013}. An ES consists of at least one named entity and one or more spatial indicators specifying its location. An ES can be identified in two ways \cite{sallaberry_semantic_2009}: as an absolute ES (ESA), it is a direct reference to a geo-locatable space (e.g. "the Plateau d'Allada"); as a relative ES (ESR), it is defined using at least one ESA and topological spatial indicators (e.g. "in Southern Benin"). These spatial indicators represent relationships, and five types of relationships are considered: orientation, distance, adjacentness, inclusion, and geometric figure that defines the union or intersection, linking at least two ES. An example of this type of ES is "near Paris".

Note that ESA and ESR integrated representation significantly reduces the ambiguities related to the identification of the right spatial footprint. Indeed, taking into account spatial indicators (e.g. "river" for "Senegal River") allows us to identify in GeoNames the right spatial footprint. To deal with distinct spatial entities with the same name (e.g. Bayonne in France and Bayonne in the United States), a disambiguation task is proposed in order to analyze the context in the textual documents \cite{kergosien_when_2015}.

In order to identify spatial entities, we apply and extend a Natural Language Process (NLP) adapted to Geographic Information Retrieval (GIR) domain. In this context, some rules (patterns) of \cite{tahrat_text2geo:_2013} have been integrated and improved to identify absolute and relative spatial entities in French (e.g. "sud-ouest de l'Arabie Saoudite" (ESR), "dans la r\'egion du Mackenzie" (ESR), "golfe de Guin\'ee" (ESA), "lac Eyre" (ESA)). These rules have been translated in English to analyze English corpora (e.g. "Willamette River" (ESA), "Indian Ocean" (ESA), "Wujiang River Basin" (ESA)) of TERRE-ISTEX project. Moreover, we propose new types of rules in order to identify Organization (for example, an "Organization is followed by an action verb"). These different rules developed with GATE1 enable to disambiguate extracted entities and then to improve named entities recognition.

\subsubsection{Thematic and temporal entities}

In order to enhance the knowledge identified in metadata and to specify the sub-domains, we apply modules in text mining to the content of the publications to extract domain vocabularies. First, we use domain semantic resources for lexical annotation. As, in our case, the thematic entities to be annotated are linked to climate change, we rely on the Agrovoc resource \cite{rajbhandari_agrovoc_2012}. Agrovoc is formalized in XML SKOS. In the indexing phase, we mark for each term the content of an article coming from Agrovoc with the terms "used for" and the generic terms, information that will be exploited in the search engine. In the long term, we aim to propose a generic approach by giving the possibility of easily integrating a new semantic domain resource formalized in XML SKOS. Also, we plan to integrate the BioTex module developed by the TETIS team in Montpellier \cite{lossio-ventura_biomedical_2016} combining statistical and linguistic approaches to extract terminology from free texts. The statistical information provides a weighting of the extracted applicant terms. However, the frequency of a term is not necessarily an appropriate selection criterion. In this context, BioTex proposes to measure the association between the words composing a term by using a measure called C-value while integrating different weights (TF-IDF, Okapi). The goal of C-value is to improve the extraction of multi-words terms that are particularly suitable for specialist fields.

For the temporal entities, we have integrated the HeidelTime processing chain \cite{strotgen_multilingual_2013} to mark calendar entities (dates and periods). HeidelTime is a free, rule-based, time-sensitive labeling system for temporal expressions, available in several languages. Regarding English, several corpora of documents (i.e., scientific articles, press) have been treated \cite{strotgen_multilingual_2013}. The evaluation of this system shows better results for the extraction and standardization of temporal expressions for English, in the context of the TempEval-2 and TempEval-3 campaigns \cite{uzzaman_semeval-2013_2013} and extended to 11 languages including French \cite{moriceau_french_2013}. HeidelTime produces annotations in the ISO-TimeML format, which distinguishes between four categories of temporal expressions: dates, times, durations and frequencies. Since our objective is to know the periods covered in the documents, we are only interested in temporal expressions with a calendar connotation.

\section{Experiments}

\subsection{First experiments}

The sociologists and geographers working in the project evaluated the spatial entities extraction process. The French corpus is composed of 4,328 words (71 spatial features and 117 organizations). The evaluations (with classical measures, i.e., precision, recall, and F-measure) have been investigated by comparing the manual extraction done by experts with the web service results. For spatial entities, we obtain a good recall (91\%) and an acceptable precision (62\%). The F-measure is 0.74. The great majority of Spatial Features (SF) are extracted but there are still some errors. The rules to identify organization are very efficient and give high precision (85\%) but the value of recall is lower (67\%). The F-measure for organization identification is 0.74. The rules for organization extraction seem well adapted to the domain but they have to be extended in order to improve the recall that remains low.

Moreover, in order to evaluate our approach of annotation of the ESA and ESR on a scientific corpus, we manually annotated 10 scientific articles in French, and 10 in English from the corpus on the 'climate change' topic. Items are randomly selected. The documents averaged 230 words and contained 39 spatial entities (i.e., ESA, ESR). We then annotated these documents with two processes: the CasEN chain, a reference in the field for the marking of named entities \cite{maurel_casen:_2011}, and ours.

We obtained good results in terms of precision, recall, and F-measure with our process (see Tables~\ref{tab:French} and~\ref{tab:English}). Nevertheless, we have issues with disambiguation of named entities (Organization and spatial entities) in English and we have to improve our process for scientific articles in English.

It is significant that the results coming from the CasEN chain are far better when processing French than English too. Assuming this difference is not directly linked to the test corpus, and besides the above disambiguation issue, we consider English linguistic specificities as an explanation. In particular, links between items of a multi-word spatial entity are seldom made explicit, neither by a common morphological feature, nor by a linking preposition.

\begin{table}[h]
 \centering
 \begin{tabular}{|l|c|c|}
 \hline
  & ESA, ESR & ESA, ESR \\
  & (TERRE-ISTEX) & (CasEN) \\
 \hline
  Precision & 100\% & 93\% \\
 \hline
  Recall & 90\% & 77\% \\
 \hline
  F-Measure & .947 & .842 \\
 \hline
 \end{tabular}
 \caption{Evaluation of spatial entity annotation on 10 articles from the French corpus}
 \label{tab:French}
\end{table}
\begin{table}[h]
 \centering
 \begin{tabular}{|l|c|c|}
 \hline
  & ESA, ESR & ESA, ESR \\
  & (TERRE-ISTEX) & (CasEN) \\
 \hline
  Precision & 90\% & 94\% \\
 \hline
  Recall & 60\% & 53\% \\
 \hline
  F-Measure & 0.72 & 0.68 \\
 \hline
 \end{tabular}
 \caption{Evaluation of spatial entity annotation on 10 articles from the English corpus}
 \label{tab:English}
\end{table}

In order to validate these initial results, we are currently working on an evaluation concerning a corpus of 600 scientific articles from the ISTEX platform, 300 written in French and 300 in English. The articles were randomly selected from the 40,000 scientific articles related to the theme of 'climate change'. In this case, on the first 140 documents manually annotated by experts working in the project, we obtain a precision of 78\%. The complete evaluation process is in progress. Producing a bigger annotated scientific corpus (with spatial entities, temporal entities and topics) is also an objective for other ISTEX research groups.

To help experts to analyze corpora, and particularly information related to territories, a Web application SISO\footnote{\url{http://geriico-demo.univ-lille3.fr/siso/}} has been developed.

\subsection{SISO Web Application to index and analyze corpora }

\begin{figure*}[t]
 \centering
 \includegraphics[width=.95\textwidth]{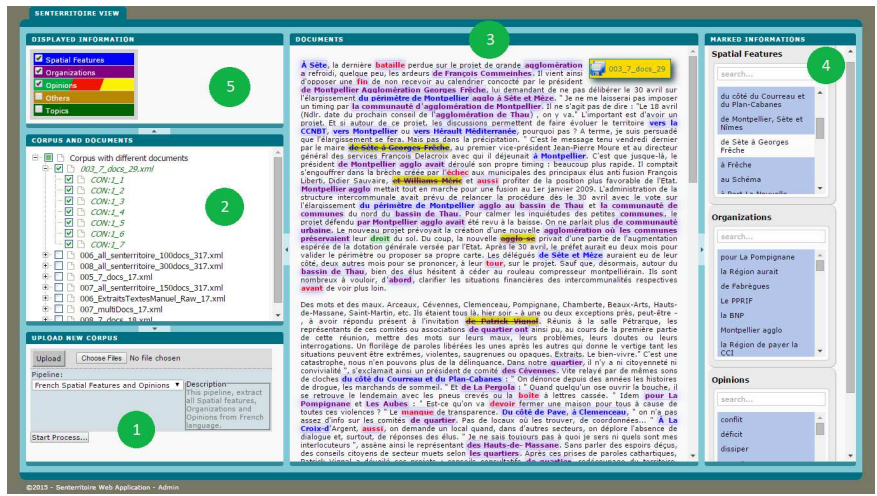}
 \caption{The Web application called SISO}
 \label{fig:SISO}
\end{figure*}

The SISO Web application (Figure~\ref{fig:SISO}) allows users to upload corpora, to index documents with specific web services in order to mark different kinds of information (spatial features, organizations, temporal features, and themes), to visualize and to correct the results, and to download validated results in XML format. More specifically, it is possible to upload corpora (frame 1), each marked corpus is saved on the server and automatically available in the web application (frame 2). After having downloaded documents, users can select the marked features (frame 5), see the results on the selected documents in frame 3. By selecting different categories from frame 5, the related marked information will be kindles in frame 3 and listed by type in frame 4. In case of finding any mistakes, users can unselect marked information (frame 4). Finally experts can export the selected corrected documents as a new corpus by clicking the top right button. The downloaded corpus, in XML MODS format, consists of selected documents with the marked information except those removed by the user.

The administration page allows users to upload, edit, and delete pipelines defined in the GATE format. It is also possible to remove processed corpora, to edit the uploaded pipeline rules and the available lexicons.

In order to provide experts with a web tool to process big data related to their domain, the TERRE-ISTEX approach was improved. The performance of the system, tested on 8,500 documents, are presented below:
\begin{itemize}
 \item Temporal entity annotation: 8,196 seconds,
 \item Agrovoc entity annotation: 1,606 seconds,
 \item Search of concept and linked concept using the offline Agrovoc ontology: 36 seconds,
 \item Spatial entities annotation (French and English): 4,940 seconds,
 \item JSON index file generation: 55 seconds.
\end{itemize}
The global process takes 16,105 seconds for processing all documents, i.e., 1.9 seconds per document. This result is very encouraging.

\section{Conclusion}

In this article, we describe a method to deal with scientific literature on climate change from three different corpora of scientific papers. One main issue was the standardizing of data. Therefore, we have developed algorithms and a unified data model. Then, we have defined an automatic process to identify information related to a territory (spatial, temporal, and thematic information) in the documents.

Up to now, the entire corpus is indexed in JSON format. We are currently working on the enrichment of the temporal entity marking chain to integrate the BioTex tool, and on the extension of tagging assessments of marked (spatial, temporal and thematic) entities in voluminous corpus. Recently, we started to index our data with the Lucene-based search engine Elasticsearch1. Elasticsearch will facilitate the test of defined work use-cases. The main goal is to help researchers analyze big data corpora, and especially those who are interested in research related to a territory.

In our future work, we plan to use machine-learning approaches in order to improve the disambiguation process of spatial entities (i.e., ES vs. Organization). More precisely, based on our previous work \cite{tahrat_text2geo:_2013}, we will propose to integrate the patterns described in section 3.1. as features in the supervised learning model based on the SVM algorithm. Then, we plan to compare our final model to the state-of-the-art, and specifically to the ISO-Space model produced to annotate Spatial Information from textual data \cite{pustejovsky_iso-space:_2017}.

\section{Acknowledgements}

This work is funded by ISTEX (https://www.istex.fr/), SONGES project (Occitanie and European Regional Development Funds), and DYNAMITEF project (CNES).

\bibliographystyle{apalike}
\bibliography{KergosienAl-lrec2018}

\end{document}